# Amplification of intense light fields by nearly free electrons

Mary Matthews[1], Felipe Morales[2], Alexander Patas[3], Albrecht Lindinger[3], Julien Gateau[1], Nicolas Berti[1], Sylvain Hermelin[1], Jerome Kasparian[1], Maria Richter[4], Timm Bredtmann[2], Olga Smirnova[2], Jean-Pierre Wolf[1], Misha Ivanov[2]

[1] *GAP, University of Geneva, 22 chemin de Pinchat, 1211 Geneva 4, Switzerland*

[2] *Max Born Institute, Max Born Strasse 2a, 12489 Berlin, Germany*

[3] *Inst. Fur Exp. Physik, Freie Universitat Berlin, Arnimallee 14, 14195 Berlin, Germany*

[4] *Departamento de Quimica, Universidad Autonoma de Madrid, 28049 Madrid, Spain*

**Light can be used to modify and control properties of media, as in the case of electromagnetically induced transparency or, more recently, for the generation of slow light or bright coherent XUV and X-ray radiation. Particularly unusual states of matter can be created by light fields with strengths comparable to the Coulomb field that binds valence electrons in atoms, leading to nearly-free electrons oscillating in the laser field and yet still loosely bound to the core [1,2]. These are known as Kramers-Henneberger states [3], a specific example of laser-dressed states [2]. Here, we demonstrate that these states arise not only in isolated atoms [4,5], but also in rare gases, at and above atmospheric pressure, where they can act as a gain medium during laser filamentation. Using shaped laser pulses, gain in these states is achieved within just a few cycles of the guided field. The corresponding lasing emission is a signature of population inversion in these states and of their stability against ionization. Our work demonstrates that these unusual states of neutral atoms can be exploited to create a general ultrafast gain mechanism during laser filamentation.**

It is often assumed that photo-ionization happens faster in more intense fields. Yet, since late 1980s, theorists have speculated that atomic states become more stable when the strength of the laser field substantially exceeds the Coulomb attraction to the ionic core [1,6-14]. The electron becomes nearly but not completely free: rapidly oscillating in the laser field, it still feels residual attraction to the core, which keeps it bound. The effective binding potential, averaged over the electron oscillations, is sketched in Figure 1(a). It has a characteristic double-well structure, the wells occur when the oscillating electron turns around near the core. The laser-modified potential also modifies the spectrum, with laser induced shifts adding to the familiar ponderomotive shift associated with nearly free electron oscillations. We refer to these states as "strongly driven laser-dressed states". In spite of many theoretical predictions, it took three decades before their existence was inferred in experiments [2,4-5], showing neutral atoms surviving laser intensities as high as $I \sim 10^{15\text{-}16}$ W/cm$^2$.

But are such unusual states really exotic? Can they also form in gases at ambient conditions, at intensities well below $10^{15\text{-}16}$ W/cm$^2$? After all, for excited electronic states bound by a few eV, the laser field overpowers the Coulomb attraction to the core at $I \sim 10^{13}\text{-}10^{14}$ W/cm$^2$. If yes, would these states manifest inside laser filaments, the self-guiding light structures created by the nonlinear medium response at $I \sim 10^{14}$ W/cm$^2$ [15]?

The formation of the KH states should modify both real [16] and imaginary [17] parts of the medium's refractive index. While their response is almost free electron-like, they do form discrete states and lead to new resonances. Crucially, at sufficiently high intensities theory predicts the emergence of population inversion in these states, relative to the lowest excited states [5,18]. If the inversion is created inside a laser filament [18], it would lead to amplification of the filament spectrum at the transition frequencies between the stabilized states.

We first confirm these expectations by directly solving the time-dependent Schroedinger equation (TDSE). We then observe these states experimentally via the emergence of absorption and stimulated emission peaks at transition wavelengths not present in the field-free atom or ion. Notably the gain takes place in neutral atoms, and we are only able to achieve gain by using shaped laser pulses, tailored to a few-cycle resolution. We also confirm theoretically that for our experimental conditions such resonances do not appear in standard filamentation models.

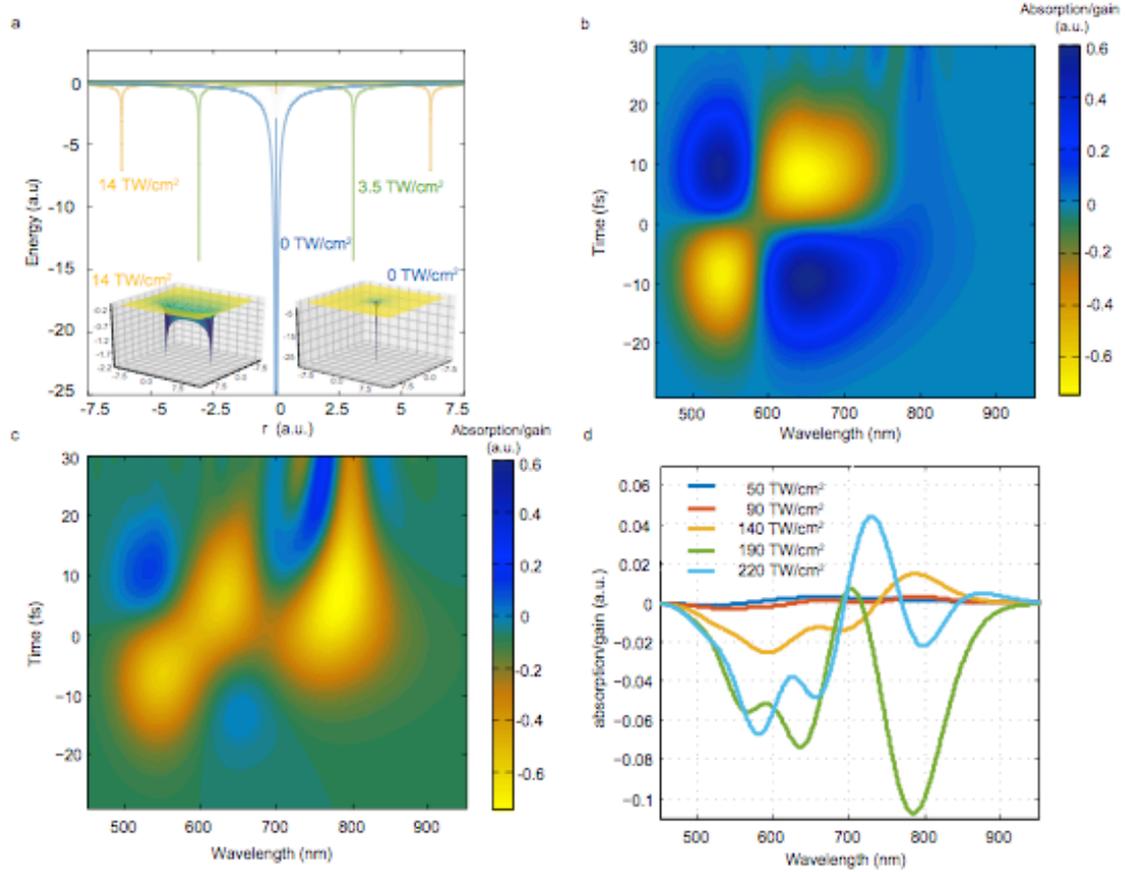

**Figure 1. Kramers Henneberger potential, and simulated absorption spectra of Ar atom dressed by a strong IR pulse**. (a) The Kramers-Henneberger potential for different pulse intensities, for 800nm, developing the characteristic double well shape. The inserts show the potential in cylindrical coordinates. (b,c) Time and frequency resolved absorption profiles, (where negative absorption signifies gain, positive – loss), for a Gabor window of T=500 a.u. and intensities of (b), 1.4 x $10^{14}$ W/cm$^2$ and (c), 1.9 x $10^{14}$ W/cm$^2$. (d) Frequency-resolved absorption during the IR pulse. Different curves correspond to different peak intensities of the dressing IR field: 0.5x $10^{14}$ W/cm$^2$ (black line), 0.9 x $10^{14}$ W/cm$^2$ (red line), 1.4 $10^{14}$ W/ cm$^2$ (green line), 1.9x$10^{14}$ W/cm$^2$ (blue line) and 2.2 x$10^{14}$ W/cm$^2$ (orange line).

Currently, lasing during laser filamentation in atmospheric gases is an active research field [15,19-28]. Recent work was also performed in low pressure Argon and Krypton [29,30], while stimulated X-ray emission has been observed from rare gas plasmas [31]. Our mechanism is novel and general for any gas: it relies on laser-dressed states in neutral atoms and uses pulse shaping to control their population and seed gain.

First, we solve the time-dependent Schrödinger equation for an Argon atom interacting with an intense, 800 nm laser field (see Methods). We use a shaped IR pulse with a sharp (~5 fs) front, which optimizes the population of the 'nearly free' laser-dressed states. Indeed, in IR fields their ionization rate maximizes at I~$10^{13}$ W/cm$^2$, (known as 'death valley'), before decreasing again at higher intensities [1,8,11-13]. Thus, the 'death valley' should be crossed quickly. The sharp front is followed by a flat top, so that the laser-dressed states are better defined. Next, we compute the linear response of the dressed atom in the *visible frequency range* to identify gain lines. To this end, the dressed atom is probed by a weak broadband (~5 fs) probe, carried at λ = 600 nm and centered in the middle of the pump pulse (t=0). The time-dependent response to the probe, Δd(t), is extracted from the full polarization d(t)=<Ψ(t)|$\hat{d}$|Ψ(t)> as described in [17]: Δd(t)=d(t)-d$_{IR}$(t). Here d$_{IR}$(t)=<Ψ$_{IR}$(t)|$\hat{d}$|Ψ$_{IR}$(t)>, Ψ(t) and Ψ$_{IR}$(t) are computed with both fields or the strong IR pump only, respectively.

The key quantity is Im[ΔD(ω)], the imaginary part of the Fourier transform of Δd(t): the negative imaginary part signifies gain, the positive signifies loss. Figures 1(b,c) show a window Fourier transform of Δd(t), using the sliding Gabor window, G$_2$(t,t0)=exp[−(t-t$_0$)$^2$/T$^2$], (T=500 a.u.), which allows us to time-resolve the emission.

Below I=$10^{14}$ W/cm$^2$, the time-dependent gain is offset by the loss, but the situation changes radically above this intensity: at I=$1.4\times10^{14}$ W/cm$^2$ gain dominates and amplification lines arise around 550-570 nm and 630-650 nm, (Figure 1(c,d)). The lines are asymmetric, more Fano-like than Lorentzian (Figure 1(d)), as expected in the presence of a strong driving field [32]. Importantly, the gain has a threshold nature and occurs intra-pulse.

Thus, theory predicts the emergence of gain at intensities I~$10^{14}$W/cm$^2$, which will manifest in the forward spectrum from only shaped (i.e. sharp rise time) laser pulses. Experimentally, we look for new, atypical absorption and emission structures with asymmetric Fano-like shapes, between 400 nm and 700 nm. Second, the population inversion should arise intra-pulse and depend on the pulse shape (rise time and duration). Third, the emission should have lasing characteristics and occur at transitions absent in the field-free atom or ion. To test these predictions we employ a pulse shaping setup [33] with a resolution down to two cycles. We use a self-phase modulated broadened and compressed Chirped Pulse Amplified (CPA) Ti:Sapphire laser in combination with a 640 pixels SLM (Spatial Light Modulator), providing 50 µJ pulses centered at 800 nm [34] (Supplementary Figure 1 (b), Methods). The pulses are focused into a chamber by a 300 mm off-axis spherical mirror, leading to a short filament,(4 mm, see Supplementary Figure 1(a) and Methods), in Ar or Kr (2-9 bar). The pulse is shaped such that it acquires the required sharp rise at the beginning of the filament, maximizing the population of the stabilized, strongly driven laser-dressed states. The pre-compensation of the desired pulse shape is achieved by acoustic shock wave optimization at the focus (see Methods). Pulse fronts ~5 fs are generated, as measured using a Spectral Phase Interferometry for Direct Electric field Reconstruction, (SPIDER).

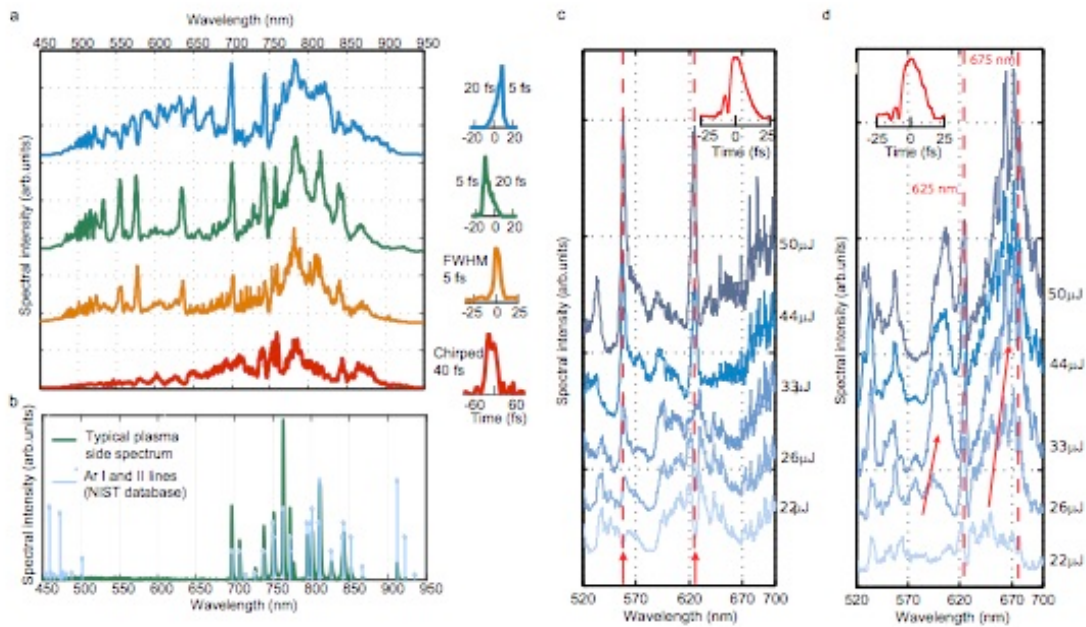

**Figure 2. Forward emission spectrum from different pulse shapes filamenting in Argon at 9 bar.** (a) shows a Gaussian pulse shape with a 40 fs duration, a 7 fs Fourier limited pulse, a sawtooth shaped pulse, where the sharp front, (5 fs), arrives first followed by a trailing 20 fs decay and the reversed sawtooth, with a 20 fs front risetime and a 5 fs decay time. The curves are offset for clarity. (b) shows the incoherent sidewards emitted spectrum from a Fourier limited, 7 fs pulse during filamentation in the same Argon cell in green, and overlaid are the corresponding Ar I and Ar II plasma recombination lines taken from the NIST database, which are not visible in the forward, coherent spectrum. (c) shows the forward emitted spectra, following filamentation in the same Argon cell, of a 10 fs rise, 5 fs plateau and 10 fs decay time, for increasing pulse energies, in steps of 5.6µJ. Two absorption features are visible at 625 nm and 560 nm which become gain features at 35 µJ and 28 µJ respectively, and a broad gain feature emerging at 600 nm, (d) shows the forward emitted spectra, following filamentation in the same Argon cell, of a 10 fs rise time, 10 fs plateau and 10 fs decay time, for increasing pulse energies. Gain features are visible at 605nm, 625nm, and 670nm. In both (c) and (d) red arrows indicate the movement of emission peaks with increasing input energy, and the dashed lines are to guide the eye to specific lasing peaks. An input white light spectrum is shown in Supplementary Figure 7(a).

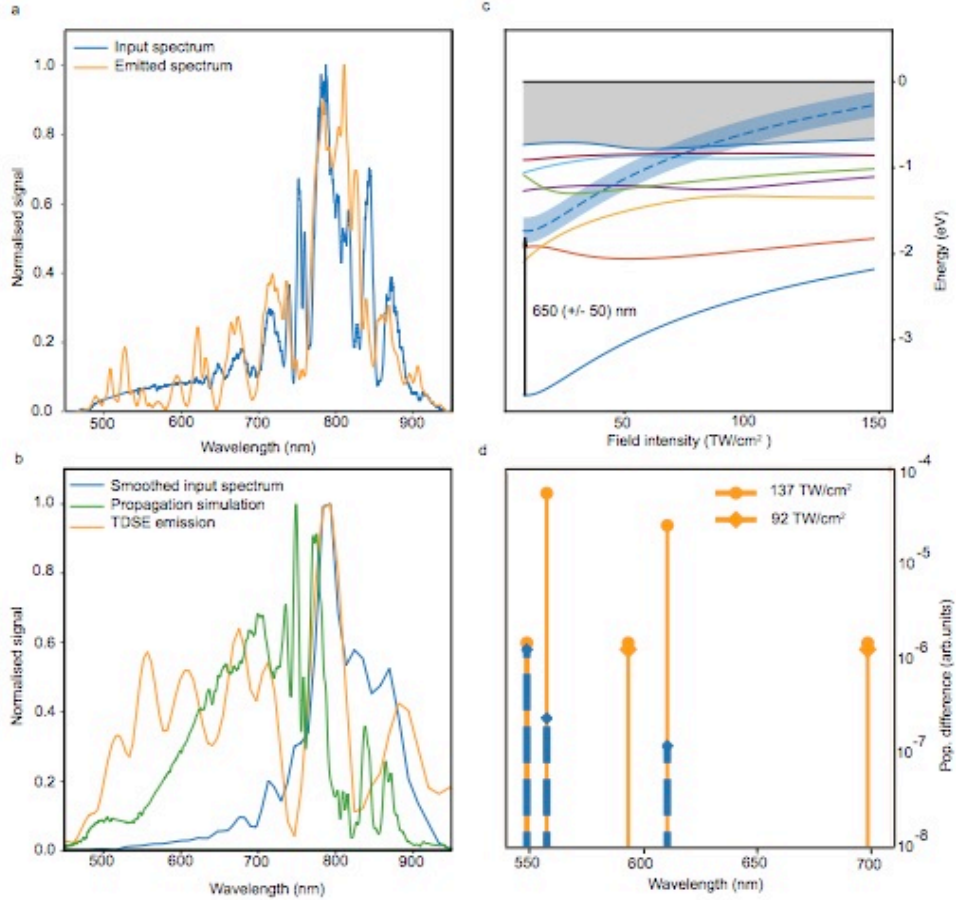

**Figure 3. A comparison between simulated and experimental emission spectra from a 10 fs rise, 10 fs plateau, 10 fs decay laser pulse shape**. (a) Experimentally measured forward emission for filamentation in Argon, at 50 μJ, showing the input, (blue) and output, (orange) spectra. (b) Theoretically calculated emission spectrum, of the strongly dressed atom for the experimental pulse at the input of the filament: input (blue), output, (orange). The green line shows the results of filamentation propagation simulations, (see Methods), without including the laser-dressed states. (c) Position of the Kramers Henneberger states as a function of laser intensity. To demonstrate the origin of the emission in the spectral region, 650 nm +/- 50 nm, the lowest excited KH state (blue) is shifted up (dashed and shaded blue): it enters the dense manifold of weakly bound states (gray region) at I~0.9 $10^{14}$W/cm². (d) The relative population difference between key field-free states, which can contribute to emission, between 500 nm and 700nm, for two different intensities: 0.92 $10^{14}$W/cm² (diamonds) and 1.37 $10^{14}$W/cm² (circles). Dashed blue lines indicate that the population of the energetically lower state is higher than that of the energetically higher state. Orange lines indicate population inversion between the states. The population is calculated at the end of the pulse.

Figures 2, 3 show the experimental results. The strongly driven laser-dressed states are best accessed using pulses with a sharp rise-time. Thus we can compare the forward emission from pulses with the same spectra, but different temporal shapes. The red line in Figure 2(a) shows supercontinuum generated inside the filament, for a smooth, 40 fs, broad Gaussian laser pulse. This standard pulse yields a typical supercontinuum spectrum in the forward direction, with no resonant lines attributable to atoms or ions. In contrast, when the pulse rise is fast, i.e. a 7 fs pulse, we observe dramatically different spectra with distinct asymmetric (Fano-like) amplification lines at 530 nm, 550 nm, 570 nm, and 625 nm, (Figure 2 (a)), as predicted by the theory. The Gaussian pulse has the seed radiation for gain or loss, but the slow rise time cannot populate the laser-driven states efficiently.

Pulse shaping control of gain is demonstrated when comparing an asymmetric triangular-like pulse (5 fs rise, 20 fs decay) against the reverse shape (20 fs rise, 5 fs fall). They have identical spectra but opposite spectral phase. The pulse with the fast rise generates strong gain lines, while the pulse with the slow rise leads to absorption at the same wavelengths. The gain lines are absent at wavelengths where no supercontinuum is present, i.e below 450 nm, as the supercontinuum acts as the lasing seed.

All the emission lines are only observed in the forward direction, indicative of emission coherent with the dressing pulse. Their divergence, measured from lateral photographs using spectral filtering, is 39 mrad in the 600 nm region, below that of the 800 nm driving pulse, (50 mrad). Their polarization is coincident with that of the driving pulse, as expected of stimulated, rather than amplified spontaneous emission. The side spectra (Figure. 2(b)) do not exhibit lines at these wavelengths, but show instead well-known Argon plasma incoherent recombination lines around 350 nm and 800 nm, (taken from the NIST database), thus the emission is not amplification of fluorescence. Above a certain threshold, the output emission intensity grows roughly linearly with the intensity of the seeding spectrum contained in the supercontinuum tail of the pulse, as expected for stimulated emission (See Supplementary Figure 2).

We now examine the dependence of gain on power and identify the lasing threshold. We use trapezoid-like pulse shapes (10 fs rise, 5 fs plateau, 10 fs fall, Figure 2 (c)), increasing the input laser energy. In Figure 2(c), the emission lines at 557 nm and 625 nm undergo absorption at lower pulse energies, but show gain when the pulse energy exceeds ~28 μJ. (For a 10 fs rise, 10 fs plateau, 10 fs fall, lasing commences at 33 μJ, Figure 2(d)). The lasing output power versus the input power, in Supplementary Figure 3, yields a lasing threshold of 1.5 GW ($I \sim 10^{14}$ W/cm$^2$). We stress that gain lines in the 610-690 nm region (highlighted resonances near 625 nm and 675 nm, Figure 2(d)), have no counterpart in the field-free spectrum of Argon, and cannot be explained by emission after the pulse.

The key role of the laser-dressed (KH) states is confirmed by the theoretical results in Figure 3. We cross-check the shape and spectrum of the trapezoidal input pulse (10 fs rise, 10 fs plateau, 10 fs rise) at the onset of filament, using numerical pulse propagation simulations (see Methods). We then use the experimental pulse in the TDSE simulations to calculate the intensity of the emitted radiation. The simulated output spectrum is normalized to the input spectrum at the 800 nm carrier wavelength, as in experiment. Figure 3 (b) shows the emergence of strong emission lines as in the experiment (Figure 3 (a)). Note these peaks emerge where Figure 1(d) shows gain. Figure 3(b) also shows that the observed lines cannot be attributed to standard non-linear effects during propagation: a simulation of laser filamentation using standard propagation models (see Methods) does not lead to any peaks in the spectral region of interest.

Finally, we focus on the spectral region between 610 nm and 690 nm. There are no field-free lines in the Argon spectrum which coincide with the observed strong amplification lines at 625 nm and near 675 nm. However, Figure 3(c) shows that transitions between the laser-dressed states (calculated in the Kramers-Henneberger frame, see Methods) do move into this region at $I \sim 0.9 \cdot 10^{14}$ W/cm$^2$. Note that Figure 3(c) does not show the overall pondermotive shift of the excited states and only demonstrates the additional shift. This shift is small compared to the pondermotive shift which reaches 6 eV at $10^{14}$ W/cm$^2$ (for λ=800 nm). Figure 3(d) shows the population difference between the field free states that move into this region at intensities around $10^{14}$ W/cm$^2$. These are the states with field free transition frequencies between 500 nm - 600 nm, which acquire population inversion at intensities around $10^{14}$ W/cm$^2$.

The lasing mechanism is not specific to Argon. Similar results were found in Krypton, see Figure 4 (and Supplementary Figure 7). The lasing transitions are at different energies than in Argon, reflecting the different atom, but also exhibit both broad and narrow gain features and asymmetric Fano-like lineshapes.

There is no direct connection between the observed resonant widths of laser dressed states, their lifetime, and pulse duration. Indeed, 1) the laser dressed states undergo ultrafast dynamics intra-pulse and 2) their positions are intensity-dependent, leading to "inhomogeneous" broadening due to spatial and temporal intensity distribution. In a 7 fs pulse, the dressed states rapidly shift with changing pulse intensity, so that resonances should broaden with increasing peak intensity (Figure 4(a), from 3 nm to 7 nm at 617 nm). For a long 'trapezoidal' pulse (10 fs rise, 40 fs plateau,10 fs rise), transition lines shift with intensity but keep their widths (~7 nm at 624 nm, and ~12 nm at 613 nm, Figure 4 (b)).

The observation of gain lines specific to the atom dressed by an intense, $I > 10^{14}$ W/cm$^2$, laser field, and absent in the spectrum of field free transitions, shows that the seemingly exotic Kramers-Henneberger states are ubiquitous even in dense (1-9 bar) gases interacting with strong laser fields. At high intensities, the laser-driven atom can become an inverted medium, inside the laser pulse, where electrons respond almost as free, yet remain bound and can be used as a multi-photon pumped gain medium during laser filamentation. Amplification at the inverted transitions between the dressed states, resulting in the emergence of gain lines during the pump pulse, can trigger additional wave-mixing processes with the strong pump, possibly leading to additional parametric gain lines in the spectrum. After the end of the pulse, coherent free induction decay can also seed lasing between the field-free states carrying population inversion. Our findings illustrate new opportunities for enhancing and controlling lasing inside laser filaments by optimizing the shape of the input laser pulse.

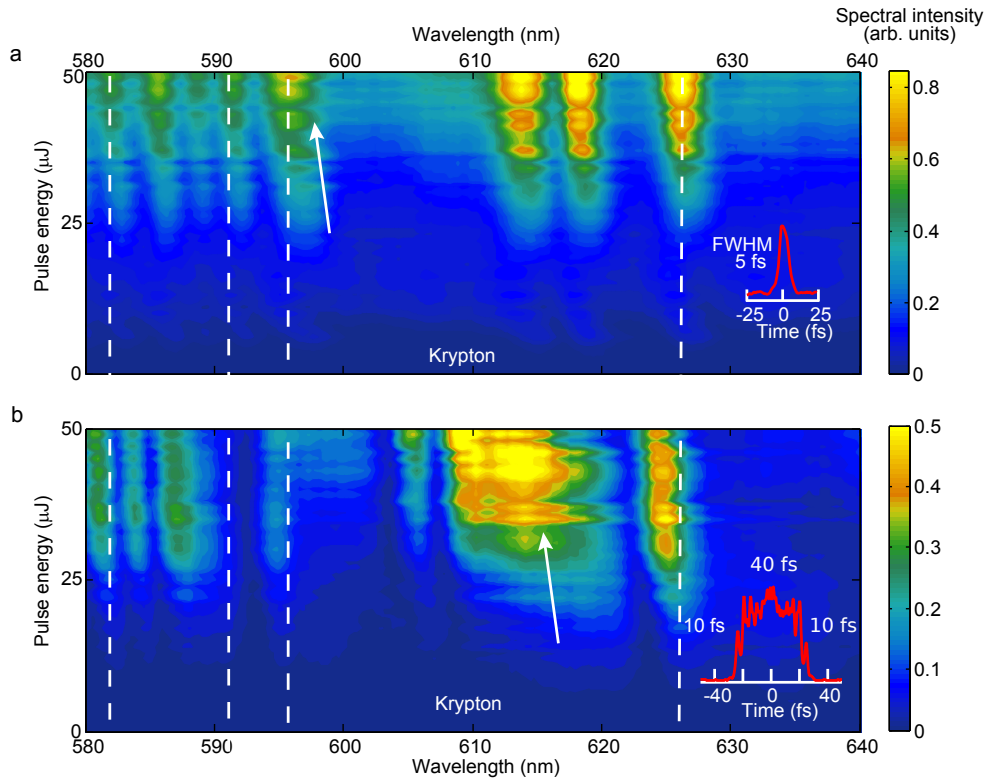

**Figure 4. The forward emission spectra of trapezoid pulse shapes, in Krypton at 9 bar, with increasing pulse energy**. (a) A Fourier limited pulse, 7 fs duration. A shift in the transition lines is observed with increasing pulse energy of 3-5 nm over 50 µJ, and at 20 µJ, the weak emission/absorption lines are strongly enhanced, and the linewidths broadened, (from 3 nm to 7 nm). (b) Trapezoid pulses with 10 fs rise time, 40 fs plateau and 10 fs decay time. Narrow and broad gain features are visible, experiencing a spectral shift of about 3-5 nm over an energy increase of 50 µJ, but with less broadening. A distinct Fano lineshape emerges at 627 nm. Dashed lines and white arrows are to guide the eye, between the Figures, to show the shift in emission wavelengths between different pulse durations and also to highlight the field dependence of the states involved. The temporal pulse shape is shown in red for each graph.

## Acknowledgements

The authors acknowledge the valuable contributions of Michel Moret, for advanced technical assistance with the experimental setup, Sebastian Courvoisier for technical assistance with graphical formatting and Ludger Woeste, for constructive advice.

J.P, J.G, S.H. acknowledge funding from SNF NCCR MUST grant. J.P and J.K acknowledge funding from ERC grant Filatmo. M.M. acknowledges funding from MHV fellowship grant number: PMPDP2-145444 and NCCR MUST Women's Postdoc Awards. M.I. acknowledges support of the DFG QUTIF grant.

## References

1. Eberly, J. H., & Kulander, K. C. Atomic stabilization by super-intense lasers. *Science* **262**, 1229-1233 (1993).

2. Eichmann, U., Nubbemeyer, T., Rottke, H., & Sandner, W. Acceleration of neutral atoms in strong short-pulse laser fields. *Nature* **461**, 1261-1264 (2009).

3. Henneberger, W. C. Perturbation method for atoms in intense light beams, *Phys. Rev. Lett.* **21**, 838 (1968).

4. Nubbemeyer, T., Gorling, K., Saenz, A., Eichmann, U. & Sandner, W. Strong-field tunneling without ionization. *Phys. Rev. Lett.* **101,** 233001 (2008).

5. Eichmann, U., Saenz, A., Eilzer, S., Nubbemeyer, T. & Sandner, W. Observing Rydberg atoms to survive intense laser fields. *Phys. Rev. Lett.* **110,** 203002 (2013).

6. Fedorov, M. V. & Movsesian, A. M. Field-induced effects of narrowing of photoelectron spectra and stabilisation of Rydberg atoms. *J. Phys. B.* **21,** L155 (1988).

7. Fedorov, M. V. & Ivanov, M. Y. Coherence and interference in a Rydberg atom in a strong laser field: excitation, ionization, and emission of light. *J. Opt. Soc. Am.* **7,** 569-573 (1990).

8. Su, Q., Eberly, J. H. & Javanainen, J. Dynamics of atomic ionization suppression and electron localization in an intense high-frequency radiation field. *Phys. Rev. Lett.* **64,** 862 (1990).

9. Pont, M., & Gavrila, M. Stabilization of atomic hydrogen in superintense, high-frequency laser fields of circular polarization. *Phys. Rev. Lett.* **65,** 2362 (1990).

10. Scrinzi, A., Elander, N. & Piraux, B. Stabilization of Rydberg atoms in superintense laser fields. *Phys. Rev. A* **48,** R2527 (1993).

11. Volkova, E. A., Popov, A. M. & Smirnova, O. V. Stabilization of atoms in a strong field and the Kramers-Henneberger approximation. *Sov. Phys. JETP.* **79,** 736-742 (1994).

12. Ivanov, M. Y., Tikhonova, O. V. & Fedorov, M. V. Semiclassical dynamics of strongly driven systems. *Phys. Rev. A* **58,** R793 (1998).

13. Morales, F., Richter, M., Patchkovskii, S. & Smirnova, O. Imaging the Kramers-Henneberger atom. *Proc. Natl. Acad. Sci. U.S.A.* **108,** 16906 (2011).

14. Fedorov, M. V., Poluektov, N. P., Popov, A. M., Tikhonova, O. V., Kharin, V. Y. & Volkova, E. A. Interference stabilization revisited. *IEEE J. Sel. Top. Quantum Electron.* **18,** 42-53 (2012).

15. Li, R., Michlberg, H. & Mysyrowicz, A. Special issue on filamentation. *J. Phys. B* **48**, Issue 9 (2015)

16. Richter, M., Patchkovskii, S., Morales, F., Smirnova, O. & Ivanov, M.. The role of the Kramers–Henneberger atom in the higher-order Kerr effect. *New J. Phys.* **15,** 083012 (2013).

17. Bredtmann, T., Chelkowski, S., Bandrauk, A. D. & Ivanov, M. XUV lasing during strong-field-assisted transient absorption in molecules. *Phys. Rev. A* **93,** 021402 (2016).

18. Bogatskaya, A. V., Volkova, E. A. & Popov, A. M. Amplification and lasing in a plasma channel formed in gases by an intense femtosecond laser pulse in the regime of interference stabilization. *Laser Physics* **26,** 015301 (2015).

19. Dogariu, A., Michael, J. B., Scully, M. O. & Miles, R. B. High-gain backward lasing in air. *Science* **331** 442-445 (2011).

20. Hemmer, P. R., Miles, R. B., Polynkin, P., Siebert, T., Sokolov, A. V., Sprangle, P. & Scully, M. O. Standoff spectroscopy via remote generation of a backward-propagating laser beam. *Proc. Natl. Acad. Sci. U.S.A.* **108,** 3130-3134 (2011).

21. Yao, J. et al. High-brightness switchable multiwavelength remote laser in air. *Phys. Rev. A* **84,** 051802(R) (2011).

22. Liu, Y., Brelet, Y., Point, G., Houard, A. & Mysyrowicz, A. Self-seeded lasing in ionized air pumped by 800 nm femtosecond laser pulses. *Opt. Express* **21,** 22791 (2013).

23. Point, G. et al. Lasing of ambient air with microjoule pulse energy pumped by a multi-terawatt infrared femtosecond laser. *Opt. Lett.* **39,** 1725-1728 (2014).

24. Malevich, P. N. et al. Ultrafast-laser-induced backward stimulated Raman scattering for tracing atmospheric gases. *Opt. Express*, **20,** 18784-18794 (2012).

25. Xu, H., Lötstedt, E., Iwasaki, A. & Yamanouchi, K. Sub-10-fs population inversion in N2+ in air lasing through multiple state coupling. *Nat. Commun.* **6,** 8347 (2015).

26. Liu, Y. et al. Recollision-induced superradiance of ionized nitrogen molecules. *Phys. Rev. Lett.* **115,** 133203 (2015).

27. Yao, J. et al. Population redistribution among multiple electronic states of molecular nitrogen ions in strong laser fields. *Phys. Rev. Lett.* **116,** 143007 (2016).

28. Luo, Q., Hosseini, A., Liu, W., & Chin, S. L. Lasing action in air driven by ultra-fast laser filamentation,

*Appl. Phys. B* **76,** 337–40 (2003).

29. Dogariu, A. & Miles, R. B. Three-photon femtosecond pumped backwards lasing in argon. *Opt. Express*, **24,** A544-A552 (2016).

30. Doussot, J., Karras, G., Billard, F., Béjot, P. & Faucher, O. Resonantly enhanced filamentation in gases. *Optica* **4,** 764-769 (2017)

31. Depresseux, A. et al. Demonstration of a circularly polarized plasma-based soft-X-ray laser. *Phys. Rev. Lett.* **115,** 083901 (2015).

32. Ott, C. et al. Lorentz meets Fano in spectral line shapes: a universal phase and its laser control. *Science* **340,** 716-720 (2013).

33. Weiner, A. M. Femtosecond pulse shaping using spatial light modulators. *Rev. Sci. Instrum.* **71,** 1929- 1960 (2000).

34. Hagemann, F., Gause, O., Woeste, L. & Siebert, T. Supercontinuum pulse shaping in the few-cycle regime. *Opt. Express* **21,** 5536 (2013).

## Methods

**Pulse shape generation.** To synthesize laser waveforms with pulse shape control down to the few cycles level, a Chirped Pulse Amplified (CPA) Ti:Sapphire laser, (780nm, 1.5mJ, 40 fs, 1 kHz, details and a diagram can be found in Supplementary Figure 1 b)) undergoes two stage filamentation in air, through loose focusing with 2 m and 1.25 m focal length mirrors. The pulse, broadened (700-900 nm) by the first filamentation stage, is re-collimated and recompressed with a pair of chirped mirrors before refocusing for the second filamentation stage, with a pair of spherical mirrors. At the exit of this second stage, the pulse spectrum spans over more than one octave (450 nm- 1 μm) and is recompressed by a chirp mirror arrangement, [34]. The final compression of higher spectral phase orders and the pulse shape control are achieved using a 4f all-reflective pulse shaper with a dual mask, 640 pixel, liquid crystal (LC) modulator. In this configuration, few cycle 5 fs pulses of up to 50 μJ can be produced, in addition to flat top, or sawtooth with sharp rise times. These are optimized using a pulse shape optimization algorithm explained below.

**Pulse shape optimization and diagnostics**. In order to compensate for dispersion arising from the chamber window and the propagation in the pressured gas before the focal point, we apply a phase detection algorithm [35], onto the Spatial Light Modulator, (SLM), to get the shortest pulse (FT limited) at the focus. The signal used for the optimization loop was the acoustic shock wave released by the plasma, representative of the free carrier density produced by the laser. Using subsequent measurements we verify this procedure leads to the desired pulse shape, at the onset of the filament (FT limited, sawtooth, flat top trapezoids). The pulse shapes are measured using a Transient-Grating FROG [34], as well as a SPIDER (Venteon), at pulse positions before and after filamentation. To measure the pulse shape within the filament, a 100 μm Al foil is placed in the filament path. The filament drills a self-adapted iris, arresting further filamentation and non-linear propagation [36], but preserving the temporal pulse shape at this distance. The remaining beam was analyzed by a SPIDER. The SPIDER traces are shown in Supplementary Figures 4, and 5.

**Pulse propagation simulations.** Numerical simulations based on unidirectional pulse propagation equation (UPPE), [37], are used to simulate the laser filamentation process and cross check the pulse shape optimization routine detailed above. The propagation simulations are first carried out up to the onset of filamentation for sample pulses and confirmed the desired experimental pulse shape. Next the same simulations were carried out throughout the full filamentation region to obtain the spectra both at the input and at the output of the filament. The numerical method and the code verification are described in detail in [38]. Briefly, the simulations are performed in cylindrically symmetric geometry, reducing the dimensionality of the problem to 2D spatial plus 1D temporal dimensions. The ionization model uses the standard Perelomov, Popov, and Terent'ev ionization rates. All standard nonlinear effects such as self-focusing, self-phase modulation, self-steepening, etc. are included, (see Supplementary Figure 8).

**Filamentation in pressured Argon cell.** A schematic of the experimental set up can be found in Supplementary Figure 1. The shaped pulses enter a pressurized chamber, (2-9 bar), containing Ar or Kr, via 5mm UVFS windows, where a 300 mm off-axis gold spherical mirror generates a filament ~4-5mm in length, before exiting the chamber through a 5mm UVFS window. Spectra from the filament and its plasma are focused in the forward and transverse directions, onto Ocean Optics fibre spectrometers (UV-Vis and NIR). An image of the filament in the transverse direction is taken by a digital camera, and the acoustic shock wave is recorded with a microphone.

**Theoretical methods.** The theoretical results in Figures 1, 3 have been obtained by propagating the TDSE

numerically, using the code described in [39]. We have used a radial box of 200.0 a. u., with a total number of radial points $n_r$ = 4000, and a radial grid spacing of 0.05 a. u. The maximum angular momentum included in the spherical harmonics expansion was $L_{max}$ = 50. The time grid had a spacing of $\Delta t$ = 0.0025 a.u. In order to remove unwanted reflections from the border of the radial box, we have placed a complex absorbing potential [40] at 32.7 a.u. before the end of the radial box. The Argon potential used was fitted to reproduce energies and dipoles of the first few one-particle states of Argon, as described in Eq. 22 of [41].

$$V_{Ar} = -\frac{1}{r}(1 + 5.4e^{-r} + 11.6e^{-3.682r})$$

To obtain the absorption spectra shown in Figure 1, we have used the technique described in [17]. The probe absorption-emission signal is proportional to:

$$S(\Omega) \propto \frac{Im[E_{PROBE}^*(\Omega)D_{PROBE}(\Omega)]}{\int d\Omega |E_{PROBE}(\Omega)|^2}$$

where $D_{PROBE}$ is the frequency resolved linear response of the IR dressed system to the probe pulse:

$$D_{PROBE}(\Omega) = \frac{1}{\Omega^2}\int dt e^{i\Omega t}[a(t) - a_{IR}(t)]$$

therefore removing the contribution of the standard nonlinear response induced by the IR.

The infrared field used in the calculations consisted of a 4 cycle $\cos^2$ turn on, followed by a 32 cycles flat top part, and a 4 cycle $\cos^2$ turn off. The carrier frequency of the dressing IR pulse is $\omega$ = 0.0569 a.u. ($\lambda$=800 nm).

The probe pulse used for extraction of the absorption spectrum of the dressed system consists of a Gaussian pulse, with central frequency $\Omega$ = 0.075942 a.u. ($\lambda$=600 nm), and a FWHM of 164 a.u.. The pulse is timed at the middle of the infrared pulse. Prior to the Fourier transform, the calculated time-dependent dipole was multiplied by a temporal mask with a flat top ending at 500 a.u. and followed by an exponential turn off with a time-constant of 200 a.u., so that the response if effectively turned off when the dressing IR pulse is over. This was done to ensure that only the dressed atom response is tracked, and that the coherent beating between the field-free states after the end of the dressing laser pulse is removed in this calculation. For the window Fourier transform with the Gabor window in Figure 1, only the Gabor window was applied, without additional exponential damping. To obtain the laser-dressed (KH) states shown in Figure 3(c), the model Argon potential was adapted to a different solver for the stationary Schroedinger equation written in cylindrical (rather than spherical) coordinates specifically for the diagonalization of the KH Hamiltonian. The approach is described in Ref.[13]. For better numerical convergence, the model potential was modified slightly while keeping the energies and the transition dipoles for all relevant states unchanged.

## Methods References

35. Wu, T. W., Tang, J., Hajj, B. & Cui, M. Phase resolved interferometric spectral modulation (PRISM) for ultrafast pulse measurement and compression. *Opt. Express* **19,** 12961-12968 (2011).

36. Xu, Z. J., Liu, W., Zhang, N., Wang, M. W., & Zhu, X. N. Effect of intensity clamping on laser ablation by intense femtosecond laser pulses. *Opt. Express* **16,** 3604-3609 (2008).

37. Kolesik, M., Moloney, J. V., & Mlejnek, M. Unidirectional optical pulse propagation equation. *Phys. Rev. Lett.* **89**, 283902 (2002).

38. Berti, N., Ettoumi, W., Hermelin S., Kasparian & J., Wolf, J. P. Non-linear Synthesis of Complex Laser Waveforms at Remote Distances. *Phys. Rev. A* **91,** 063833 (2015).

39. Muller, H. G. Simple, accurate, and efficient implementation of 1-electron atomic time-dependent Schrödinger equation in spherical coordinates. *Comput. Phys. Commun*. **199**, 153-169, (2016).

40. Manolopoulos D. E. Derivation and reflection properties of a transmission-free absorbing potential. *J. Chem. Phys.* **117,** 9552–9559 (2002).

41. Muller, H. G. Numerical simulation of high-order above-threshold-ionization enhancement in argon. *Phys. Rev. A* **60,** 1341, (1999).

# Amplification of intense light fields by nearly free electrons

# Supplementary Material

Mary Matthews[1], Felipe Morales[2], Alexander Patas[3], Albrecht Lindinger[3], Julien Gateau[1], Nicolas Berti[1], Sylvain Hermelin[1], Jerome Kasparian[1], Maria Richter[4], Timm Bredtmann[2], Olga Smirnova[2], Jean-Pierre Wolf[1], Misha Ivanov[2]

[1] GAP, University of Geneva, 22 chemin de Pinchat, 1211 Geneva 4, Switzerland

[2] Max Born Institute, Max Born Strasse 2a, 12489 Berlin, Germany

[3] Inst. Fur Exp. Physik, Freie Universitat Berlin, Arnimallee 14, 14195 Berlin, Germany

[4] Departamento de Quimica, Universidad Autonoma de Madrid, 28049 Madrid, Spain

**Contents:**

Supplementary Figures 1 – 8

Supplementary Text: Note on control of lasing

Supplementary Figure 1

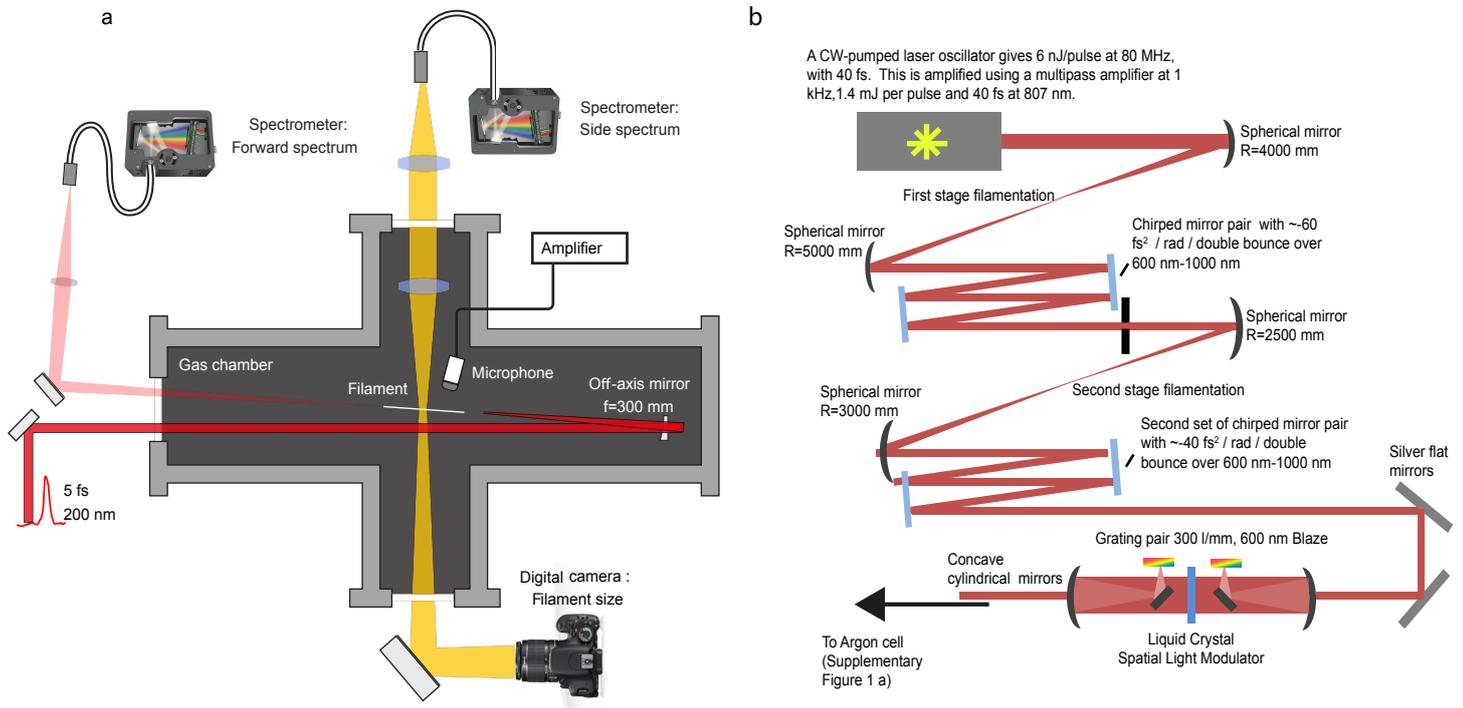

**Supplementary Figure 1. Illustrations of the laser pulse shaping system and the filamentation set up within a pressured Argon cell.** (a) shows the pressured Argon cell, where a shaped pulse is focused inside with either a spherical focusing mirror or an off axis parabolic mirror, (f=30 cm), and forward and side spectra measurement are taken with Ocean Optics HR4000 spectrometers. Filament length is monitored with a Nikon camera while the shock wave due to the plasma creation is monitored with a microphone. The mirrors are coated for broadband reflectivity, and the gold off-axis parabolic mirror has a focal length of 30 cm. We use ultra violet fused silica windows and lenses, for the cell and focusing into the spectrometers. Details of the broadband laser and pulse shaping and compression system can be found in Methods Refs. [36] and [38]. (b) shows an illustration of laser system used to generate pulse shapes: we use a frequency-doubled Nd:Vanadate laser (Verdi V5, Coherent) to pump a Ti:Sapphire oscillator (Femtosource Compact, Femtolasers) to produce 6 nJ centered at 805 nm, 90 nm bandwidth at 80 MHz. This seeds a multi-pass Ti:Sapphire chirped pulse amplification (Odin C, Quantronix) at 1 kHz, pumped with a nanosecond frequency-doubled Nd:YLF. The amplified pulses are at 807 nm with a bandwidth of 46 nm, energy ranging from 0.4 to 1.4 mJ and pulse durations of sub 40 fs. A first stage of filamentation in air, using a 2.5 m spherical mirror, gives moderate broadening with a pulse energy of 430 μJ. Following collimation and compression, (two double bounces on a GVD-oscillation compensated chirped mirror pair (Layertec)), a second stage of filamentation is performed, with a 1.25m focusing spherical mirror. Further compression is then completed with a chirped mirror pair (Layertec).

Supplementary Figure 2

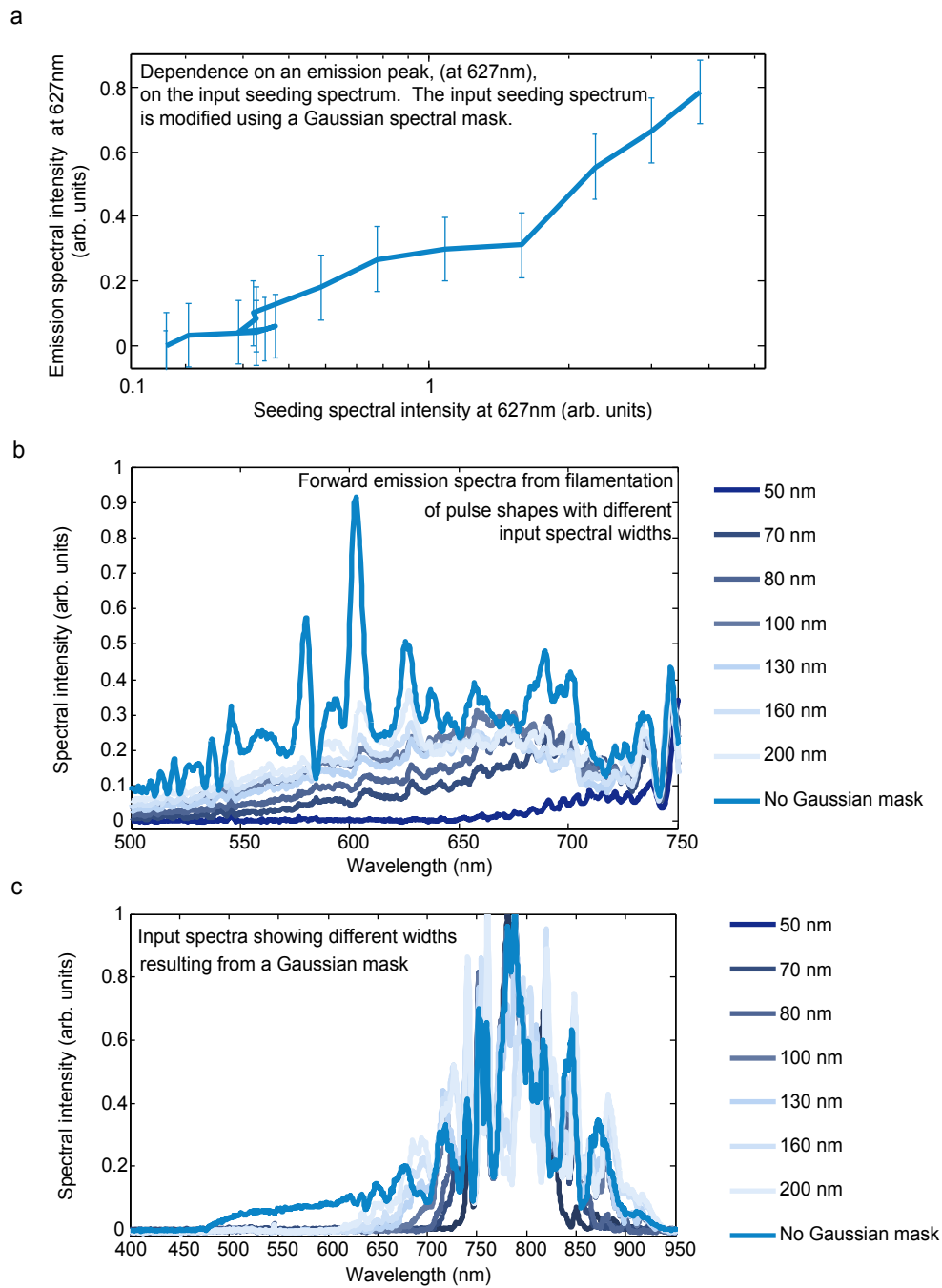

**Supplementary Figure 2  The dependence of the spectral intensity of a resonant lasing peak on the intensity of the input seeding spectrum.**  (a) shows the dependence of a lasing peak at 625 nm on the intensity of the seeding spectrum at 624 nm - 626 nm  while (b) shows the observed forward emission spectra.  The seeding spectral intensity is controlled by applying a series of Gaussian spectral masks to the input spectrum, to gradually reduce the supercontinuum tail.  The Gaussian spectral masks are shown in (c).  The rise time, and the pulse duration remain short, (sub 18 fs), and the overall pulse energy is kept constant.  There is a clear dependence on the seed radiation, indicative of stimulated rather than amplified spontaneous emission.

Supplementary Figure 3

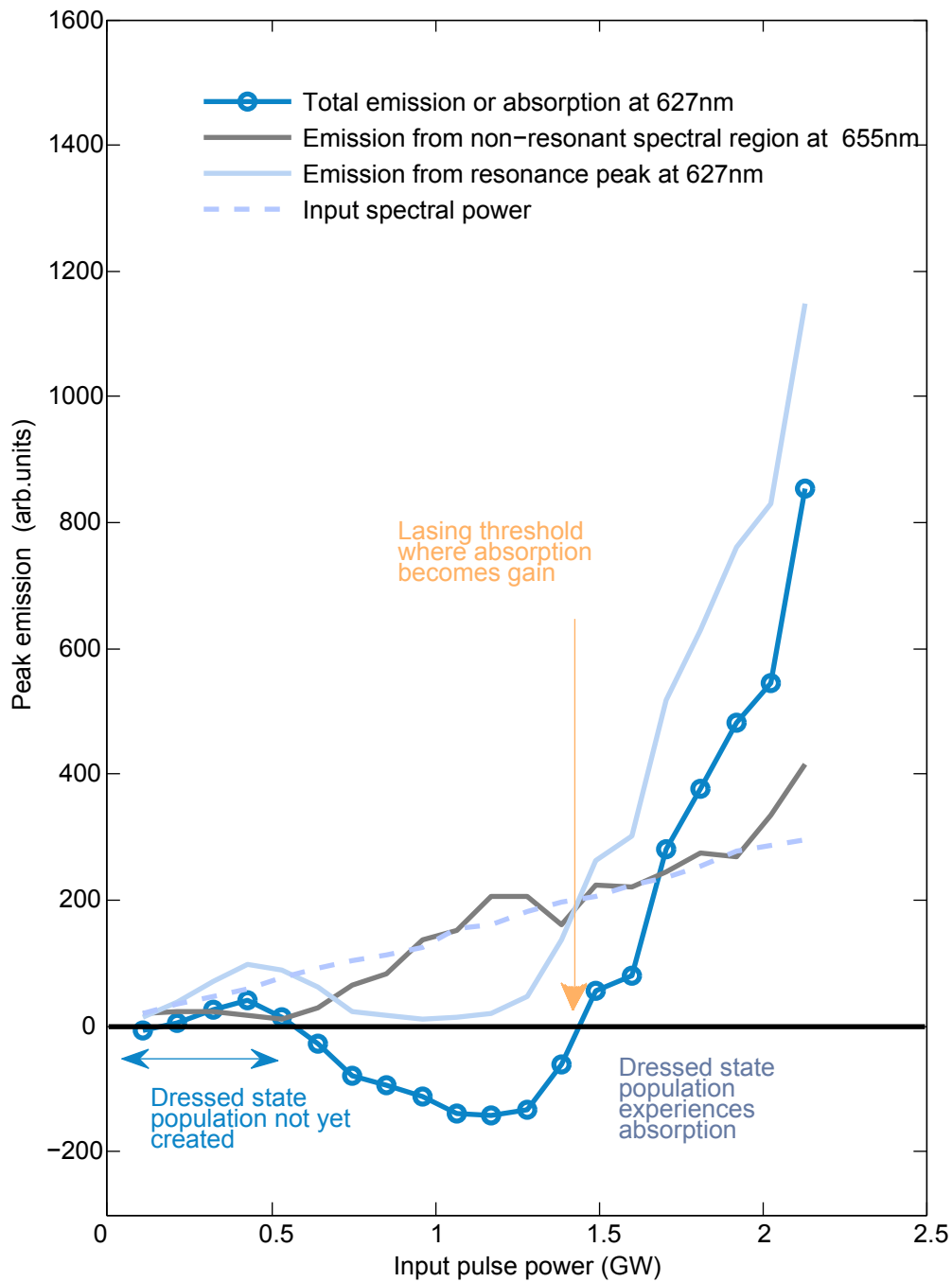

**Supplementary Figure 3. The dependence of spectral intensity of a resonant lasing peak on the input power of the shaped pulse**. We plot the output spectral power of the emission peak, (at 625 nm) as well as an adjacent region of supercontinuum (at 655 nm) against the input dressing pulse power, for the trapezoid pulse, shaped as 10 fs – 5 fs - 10 fs. For the resonant peak at 625 nm and a non-resonant region at 655 nm, the resonant peak increases with a gradient of six times the rate of the non-resonant spectral region. Subtracting the input spectrum, we observe an absorption region followed by a lasing threshold point and a gain region.

Supplementary Figure 4

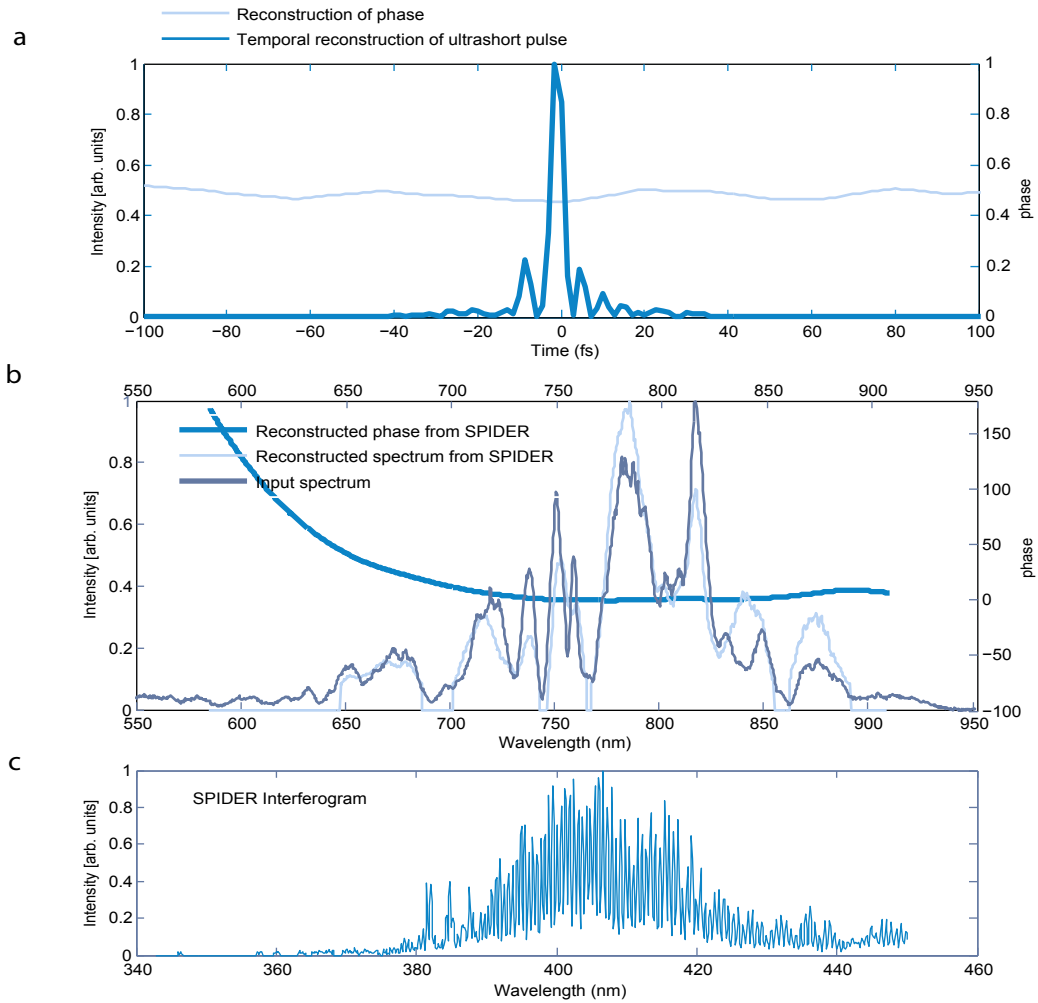

**Supplementary Figure 4. SPIDER measurements and pulse reconstruction of an ultrashort pulse.** The SPIDER was taken by placing a piece of Aluminium foil at the centre of the laser filament inside the Argon chamber. A hole was drilled by the pulse at high intensity, which allowed the passage of the central region of the pulse, removing the photon bath around the filament and reducing the intensity, before input into the Venteon SPIDER. Tests were performed both outside and inside the cell, with known chirp added using the pulse shaper, to confirm the accuracy of the reconstructions. (a) shows the pulse temporal reconstruction and phase, (b) shows the spectral reconstruction and (c) shows the raw interferogram. The measurements are at the limit of the bandwidth of the crystal, and the contribution of short wavelengths is underestimated due to the comparatively lower power of supercontinuum tail.

Supplementary Figure 5

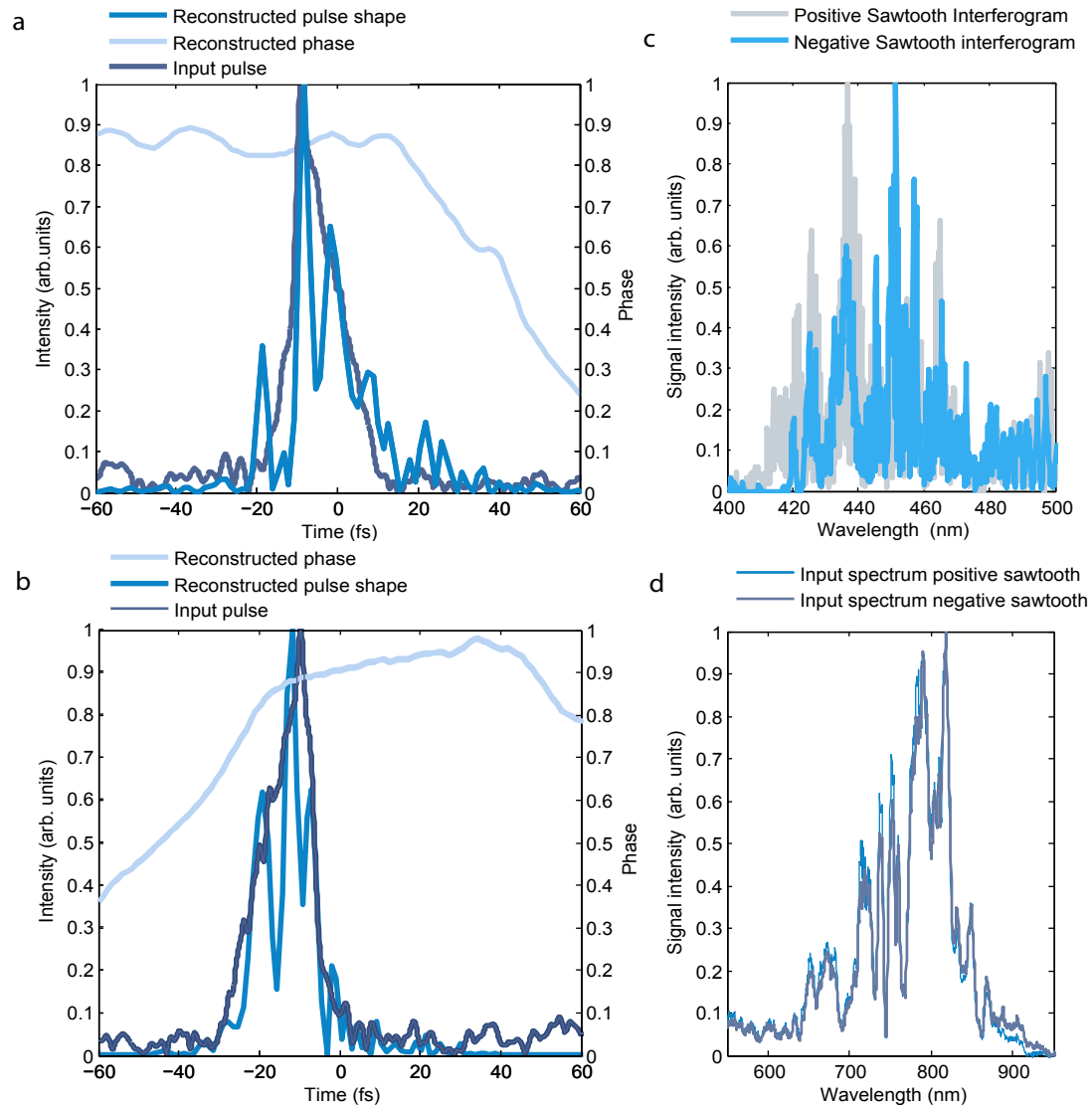

**Supplementary Figure 5. SPIDER measurements and pulse reconstruction of asymmetric triangular pulses, with a fast rise and a slow decay or vice versa.** The SPIDER interferograms are taken at the position of the filament, by piercing a sheet of aluminum foil to take only the central region of the pulse in space and remove the photon bath. The reduced energy of the central region was then fed into a Venteon SPIDER and an interferogram trace was recorded. (a) shows a positive sawtooth, with a fast rise time, with the reconstructed phase and temporal shape. The input pulse shape, (without precompensation), is overlaid in dark blue. (b) shows the negative sawtooth reconstruction and input pulse. (c) shows the corresponding interferograms, and (d) shows the input spectrum. There is good agreement, and we retrieve equal but opposite temporal phases. The measurements are at the limit of the bandwidth of the crystal, and the contribution of short wavelengths is underestimated due to the comparatively lower power of supercontinuum tail.

Supplementary Figure 6

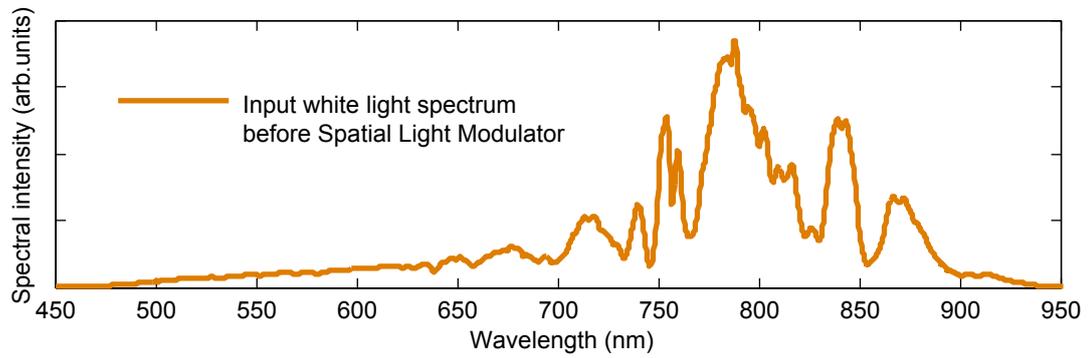

**Supplementary Figure 6. White light input spectrum without pulse shaping, following two stage filamentation.** The spectrum is centred around 780 nm.

Supplementary Figure 7

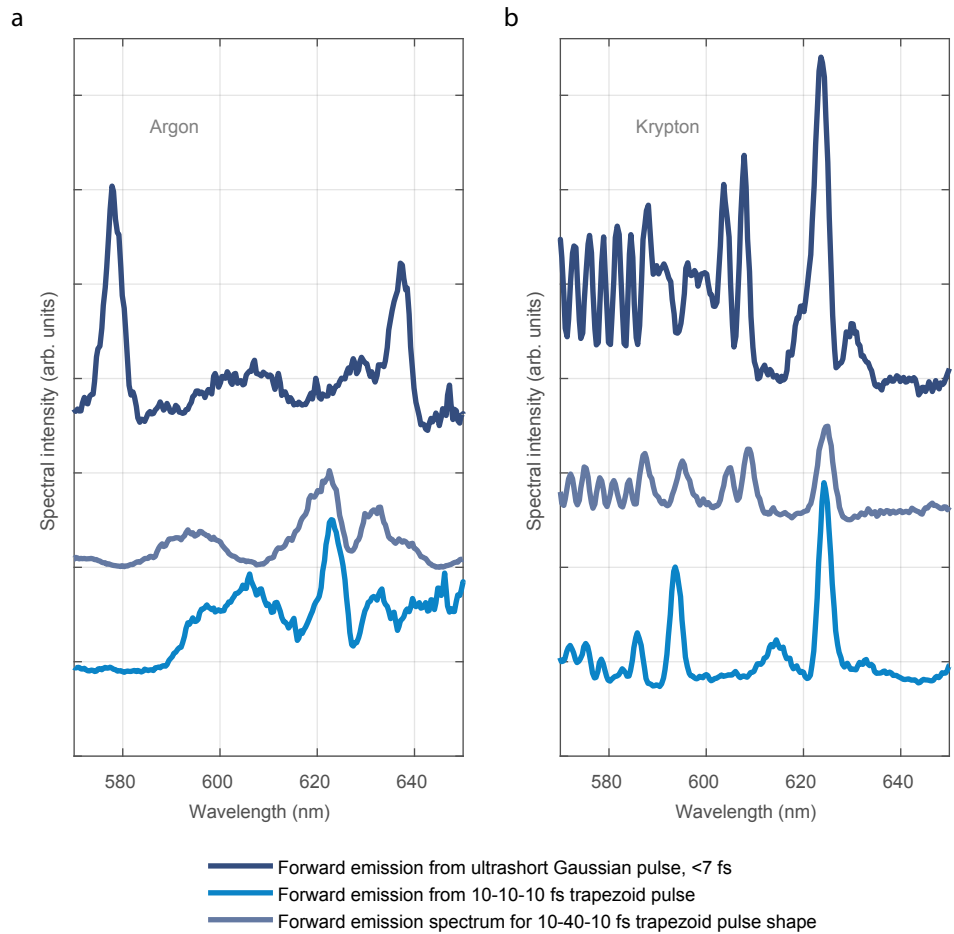

**Supplementary Figure 7. Direct Comparison between Argon and Krypton gases for an ultrashort, 7 fs pulse.** (a) shows Argon and (b), Krypton. The strong emission features are located between 580nm and 640nm, but at different locations, reflecting the different transitions in the laser dressed atom.

Supplementary Figure 8

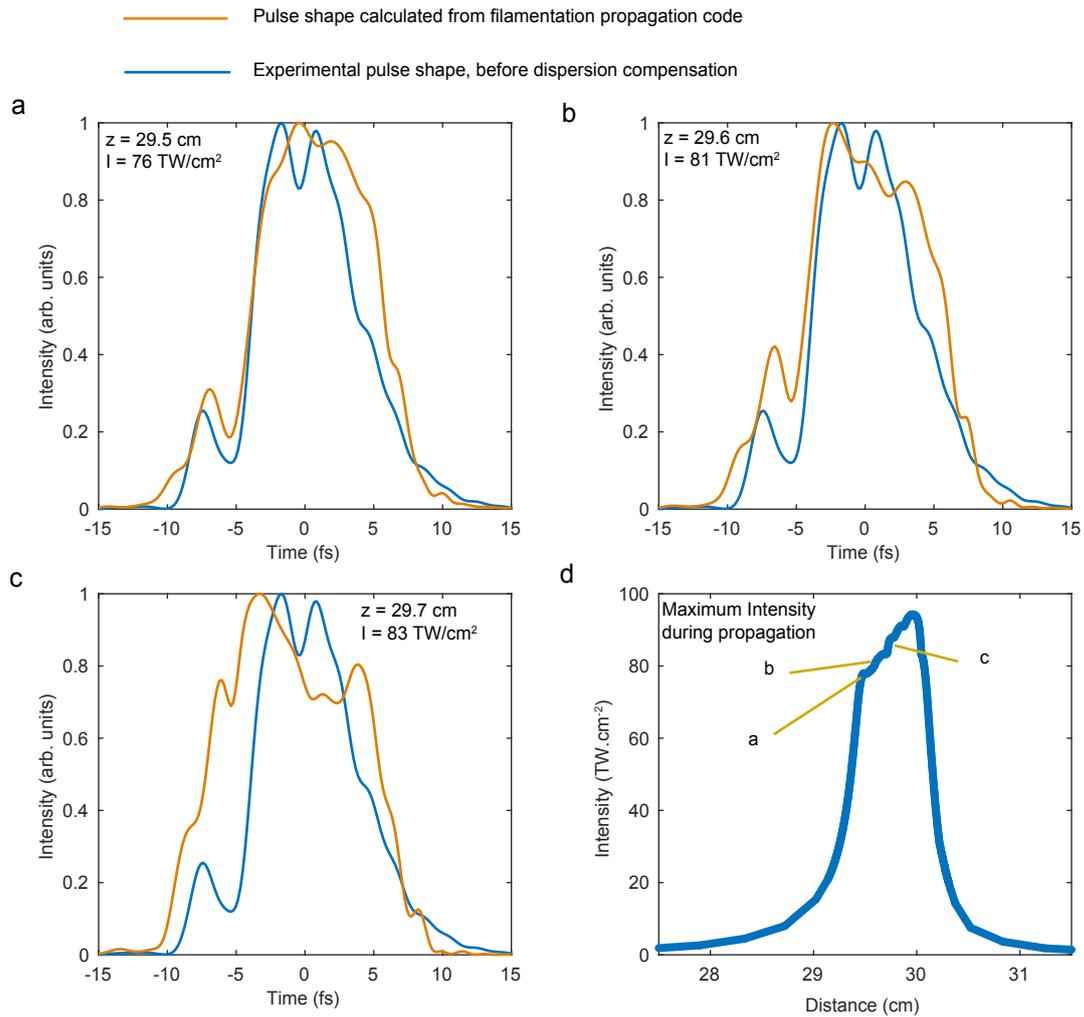

**Supplementary Figure 8. Propagation simulations of laser filamentation for a pulse shape with a 10 fs rise- 10 fs plateau – 10 fs decay.** (a), (b) and (c) show the pulse temporal form at three different longitudinal positions at the onset of filamentation. (see Methods for calculation details). Overlaid is the desired pulse temporal shape. (d) shows the intensity change across the filamentation region of ~5mm. The calculated filamentation spectrum is shown in Figure 3 (b) of the text.

# Supplementary Text 1

**A note on control of lasing.** We have already demonstrated the sensitivity of lasing to the seed spectrum. We show that by altering the pulse shape duration, we change the relative populations of the dressed states. In Argon, a trapezoid with a 10fs plateau leads to the emergence of a resonance at 600nm, not present in the 5 fs plateau pulse, (see Figure 2). In Krypton, Figure 4, moving from a short, ~5fs pulse to a 40 fs trapezoid leads to a distinctly different spectrum, with new transitions appearing (606nm) and others no longer visible, (591nm and 618nm) . Thus, the dependence on the seeding spectrum (i.e. the supercontinuum tail of the input spectrum) and on the dressing pulse shape allows us to enhance selected lasing wavelengths while suppressing others.